\documentclass[letterpaper]{JHEP3}
\usepackage{epsf,epsfig}
\usepackage{subfigure}
\usepackage{graphicx}
\usepackage{amsmath}
\usepackage{bbold}
\usepackage{slashed}

\newcommand\rep\mathbf

\newcommand\fold\otimes
\def\etmiss{E\!\!\!\!\slash_{T}}

\newcommand\dashlength5 
\preprint{MADPH--12-1579}

\title{\vspace*{.75in}
Uncovering the Charming Higgs at the LHC}

\author{Ian Lewis$^{a,b}$ and Jared Schmitthenner$^b$\thanks{ ilewis@bnl.gov,\ schmitthenne@wisc.edu}\\
{$^a$\it Department of Physics, Brookhaven National Laboratory, Upton, NY 11973, U.S.A.}\\
{$^b$\it Department of Physics, University of Wisconsin, Madison, WI 53706, U.S.A.} }

\abstract{We study the observability of the Higgs boson in the ``charming Higgs" model.  In this model the Higgs boson primarily undergoes a cascade decay to four charm quarks via light intermediate pseudoscalars. Such a decay allows the Higgs boson to escape the most stringent LEP bounds on the Standard Model Higgs boson mass.  If the light pseudoscalars are sufficiently light they become highly boosted and their decay products collimated into jets.  We show that by using jet substructure techniques, this model is potentially observable at the LHC.  For a Higgs boson mass of $100$~GeV and light pseudoscalar mass of $12$~GeV, we find a signal significance of $3.8\sigma$ with a luminosity of $30$ fb$^{-1}$ and that a $5\sigma$ significance can be obtained with $50$~fb$^{-1}$ of luminosity at the $14$ TeV LHC.%
 }
\keywords{Higgs Boson, Subjets, LHC}

\begin{document}
\section{Introduction}

The mechanism of electroweak symmetry breaking (EWSB) is one of the central unanswered questions of particle physics.  While the Standard Model (SM) with one Higgs doublet provides a simple EWSB mechanism that is so far in agreement with experiment, LEP has already excluded a SM Higgs with mass below $114.4$~GeV~\cite{Barate:2003sz}.  Also, the ATLAS and CMS experiments at the Large Hadron Collider (LHC) are quickly excluding the SM Higgs.  At~$95\%$ confidence level, the ATLAS experiment currently rules out the mass ranges $110-117.5$, $118.5-122.5$, and $129-539$ GeV~\cite{ATLAS}, and CMS excludes $129-525$ GeV~\cite{CMS}. As the limits on the SM Higgs boson tighten, we are motivated to investigate other possible mechanisms for EWSB.  Two such examples are supersymmetry and technicolor~\cite{Weinberg:1975gm}.  

One may also note that electroweak precision constraints prefer a SM Higgs with mass at or slightly below $100$~GeV~\cite{:2010zz}, beneath the LEP direct detection bounds.  Hence, a particularly interesting class of models are those that contain a SM-like Higgs boson that escapes the strictest LEP bounds and has mass nearer that preferred by precision measurements.  This can be accomplished in theories with a SM-like Higgs boson that preferentially undergoes a cascade decay into light SM particles.  In particular, if the Higgs boson dominantly decays via a cascade $h\rightarrow 2\eta\rightarrow 4X$, where $\eta$ is a pseudoscalar and $X$ are light SM particles, the LEP bounds can be significantly relaxed~\cite{Dermisek:2005ar,Dobrescu:2000yn}.

There has been much recent work on detecting a Higgs boson decaying to two pseudoscalars and the pseudoscalar subsequently decaying to $2\gamma$, $2\tau$, $2b$, $2\mu$ and $2g$~\cite{Chang:2006bw,Falkowski:2010hi,Chen:2010wk}.   If the pseudoscalars preferentially decay to two $b$'s, the Higgs mass bounds from LEP are only mildly reduced to $110$~GeV~\cite{Schael:2006cr}. Hence, to evade the most stringent LEP bounds, many models of this type require the mass of the pseudoscalar to be less than the bottom quark threshold, $2m_b$, eliminating the $\eta\rightarrow b\bar{b}$ decay.  However, one example, the so-called ``charming Higgs" model~\cite{Bellazzini:2009kw}, has matter representations that suppress the couplings of the pseudoscalar to down type quarks and charged leptons. Consequently, the pseudoscalar decay to charm quarks is dominant over the decay to bottom quarks or $\tau$ leptons, even for pseudoscalar masses above the bottom quark threshold.  In such a scenario, the Higgs boson can still have a mass around $m_Z$ and avoid the most stringent LEP bounds.

In this paper we study the Higgs production in association with a $W$ at the LHC with a subsequent Higgs decay $h\rightarrow2\eta\rightarrow4c$.  Specifically, we consider the case where the Higgs boson mass, $m_h$, is below the LEP bounds and the $\eta$ mass, $m_\eta$, is above the bottom quark threshold, $m_\eta> 2m_b$.  Hadronically decaying Higgs bosons are difficult to observe at the LHC due to the large QCD backgrounds.  We therefore consider the scenario where $m_\eta\ll m_h$.  In this case the $\eta$s become highly boosted and their decay products are collimated into single jets.   
Our signal jets then originate from color singlet particles decaying to two quarks while the QCD background jets originate from massless colored partons.  Hence, the signal jets have a mass scale and substructure distinct from the background.
 Using the techniques developed in Ref.~\cite{Butterworth:2008iy}, we decompose the jets into subjets and perform a substructure analysis.

The paper is structured as follows.  In Section~\ref{Model.SEC} we outline the simplified model we utilize.  We study the observability of the charming Higgs associated production with a vector boson via a complete signal and background analysis in Section~\ref{Numerical.SEC}.  In Section~\ref{Subjet.SEC}, we demonstrate that a subjet analysis improves the observability of the charming Higgs at the LHC.  Finally, in Section~\ref{Conc.SEC} we summarize our results and conclude.

\section{Simplified Model}
\label{Model.SEC}
In the models of interest~\cite{Bellazzini:2009kw,Bellazzini:2009xt}, the Higgs boson arises as a pseudo-Goldstone boson (pGB) of an approximate global symmetry.  The symmetry breaking pattern of these models gaurantees an additional light SM singlet pseudoscalar, $\eta$.  The pseudoscalar then has derivative couplings to the SM-like Higgs boson, $h$:
\begin{eqnarray}
\mathcal{L}\approx -h(\partial_\mu \eta)^2\frac{v_{EW}}{\sqrt{2} f^2}\left(1-\frac{v^2_{EW}}{f^2}\right)^{-1/2}
\end{eqnarray}
where $v_{EW}=174$~GeV is the electroweak breaking scale, and $f$ is a global symmetry breaking scale.  For $f$ values on the order of the electroweak breaking scale, $h$ will dominantly decay to two $\eta$'s.  For example, if $f=350-400$~GeV and the Higgs boson mass, $m_h$, is around the $Z$-mass, the branching ratio of $h$ to $2\eta$ is $80-90\%$ and $h$ to $b\bar{b}$ is $10-20\%$, consistent with the LEP bounds~\cite{Bellazzini:2009kw,Bellazzini:2009xt}.

The couplings of the pseudoscalar to a SM fermion are of the form
\begin{eqnarray}
iy_{f}\eta \bar{f}\gamma_5 f.
\label{etaFF.eq}
\end{eqnarray}  
Typically, $\eta$ couples most strongly to top and bottom quarks.  Hence, for $m_\eta>2m_b$, the $\eta$ predominantly decays into $b\bar{b}$ and the LEP bound is only slightly alleviated to $m_h>110$~GeV~\cite{Schael:2006cr}.
  However, in the so-called ``charming Higgs" model~\cite{Bellazzini:2009xt}, the $\eta$ coupling to bottom quarks is suppressed by higher order operators.  Also, $\eta$ couples to $\tau$s through the mixing of $\tau$ with a heavy partner, whereas the coupling to charm quark does not suffer from these suppressions.  Hence, the dominant decay mode is then $\eta\rightarrow c\bar{c}$ for all values of $m_\eta$.   

For simplicity and clarity, throughout the rest of this paper we take the branching ratios of the Higgs boson BR$(h\rightarrow\eta\eta)=1$ and pseudoscalar BR$(\eta\rightarrow c\bar{c})=1$.  For smaller branching ratios, the signal cross sections can be simply scaled.  

\section{Signal and Background}
\label{Numerical.SEC}
In the context of the simplified model in Sec.~\ref{Model.SEC}, we study the production of a Higgs boson in association with a $W$ boson, $pp\rightarrow Wh$, with the subsequent decay $h\rightarrow 2\eta\rightarrow 4c$.  Additionally, we consider events where the vector boson decays to leptons, which allows our signal events to be more easily triggered and avoids pure QCD backgrounds.  We allow the $W$ to decay to both electrons and muons. The associated production with a $Z$-boson is also possible; however, the branching ratio of $Z$ to leptons is less than $W$ to leptons and so we focus on purely $Wh$ production.  Similar studies have been performed for $h\rightarrow 4g$ and $h\rightarrow 4b$~\cite{Falkowski:2010hi,Chen:2010wk}.

 All signal and background events are generated using MadGraph5~\cite{Alwall:2011uj} and then showered using Herwig~6.5.10~\cite{Corcella:2000bw}.  The charming Higgs model is incorporated into MadGraph using FeynRules~v1.6~\cite{Christensen:2008py}.  Jets are clustered using FastJet~v2.4.2~\cite{Cacciari:2006sm} and the $k_T$ algorithm~\cite{Catani:1993hr} with a radius $R=0.5$.

For the purposes of this study, we consider a Higgs mass of $100$~GeV and an $\eta$ mass of $12$~GeV.  With this parameter choice the $\eta$s will be highly boosted and their individual decay products will be indistinguishable.  Hence, our signal consists of a lepton, two jets, and missing energy.  After showering and clustering, we require at least two jets to pass the basic acceptance cuts
\begin{eqnarray}
p^j_T>30~{\rm GeV}~~~~~~|y^j|<2.5,\label{cuts1J.EQ}
\end{eqnarray}
where $p_T$ and $y$ are transverse momentum and rapidity, respectively.  Additionally, to trigger on the signal we apply the lepton and missing transverse energy cuts
\begin{eqnarray}
&&p^\ell_T>30~{\rm GeV}~~~~~ |y^\ell|<2.5~~~~~~ \etmiss>25~{\rm GeV}.\label{cuts1L.EQ}
\end{eqnarray}
The signal cross section after these cuts is shown in the second column of Table~\ref{xsect.TAB}.

\begin{table}[tb]
\caption{Cross section for signal ($Wh$) and background, signal to background ratio, and signal significance with consecutive cuts at the 14 TeV LHC.  Negligible backgrounds are indicated by --.}
\label{xsect.TAB}
\begin{center}
\begin{tabular}{|l|c|c|c|c||c|}  \hline
$\sigma$(fb)                  & Cuts Eq.~(\ref{cuts1J.EQ})+(\ref{cuts1L.EQ})  & +~Eq.~(\ref{cuts2.EQ})  & +~$n^{\rm pass}_j=2$     & +~Eq.~(\ref{cuts3.EQ})   & +subjet cuts\\ \hline
 \hline
$Wh$ & 84 & 27 & 22 & 3.3 & 1.1 \\ \hline\hline
$Wjj$ & $1.4\times 10^6$ & $5.3\times10^4$  & $4.4\times10^4$ & 450 & 1.2 \\ \hline
$WW$ & $1.7\times 10^3$ & $41$ & $31$ & $0.91$ & $1.3\times 10^{-2}$ \\ \hline
$WZ$ & $470$ & $50$ & $42$ & $13$ & $1.2\times10^{-2}$ \\ \hline
$tq$ & $7.1\times 10^3$ & $170$ & $140$ & $0.28$ & -- \\ \hline
$tW$ & $3.3\times 10^3$ & $150$  & $76$ & $3.5$ & $3.3\times10^{-2}$ \\ \hline
$tbW$ & $2.3\times10^5$ & $7.5\times 10^3$ & $1.5\times10^3$ & 78 & -- \\ \hline\hline
~~~$t\bar{t}$ & $4.3\times 10^4$ & $1.4\times 10^3$ & $96$ & 26 & -- \\ \hline\hline
$S/B$ & $5.1\times10^{-5}$ & $4.4\times10^{-4}$ & $4.9\times10^{-4}$ & $6.1\times10^{-3}$ & 0.85 \\ \hline
$S/\sqrt{S+B}$ ($30$ fb$^{-1}$) & 0.36  & 0.60 & 0.57 & 0.78 & 3.8 \\ \hline
$S/\sqrt{S+B}$ ($100$ fb$^{-1}$) & 0.66 & 1.1 & 1.0 & 1.4 & 7.0 \\ \hline
\end{tabular}
\end{center}
\end{table}

\begin{figure}[tb]
\centering
\subfigure[]{
	\label{mjj.FIG}
	\includegraphics[width=0.31\textwidth,clip,angle=-90]{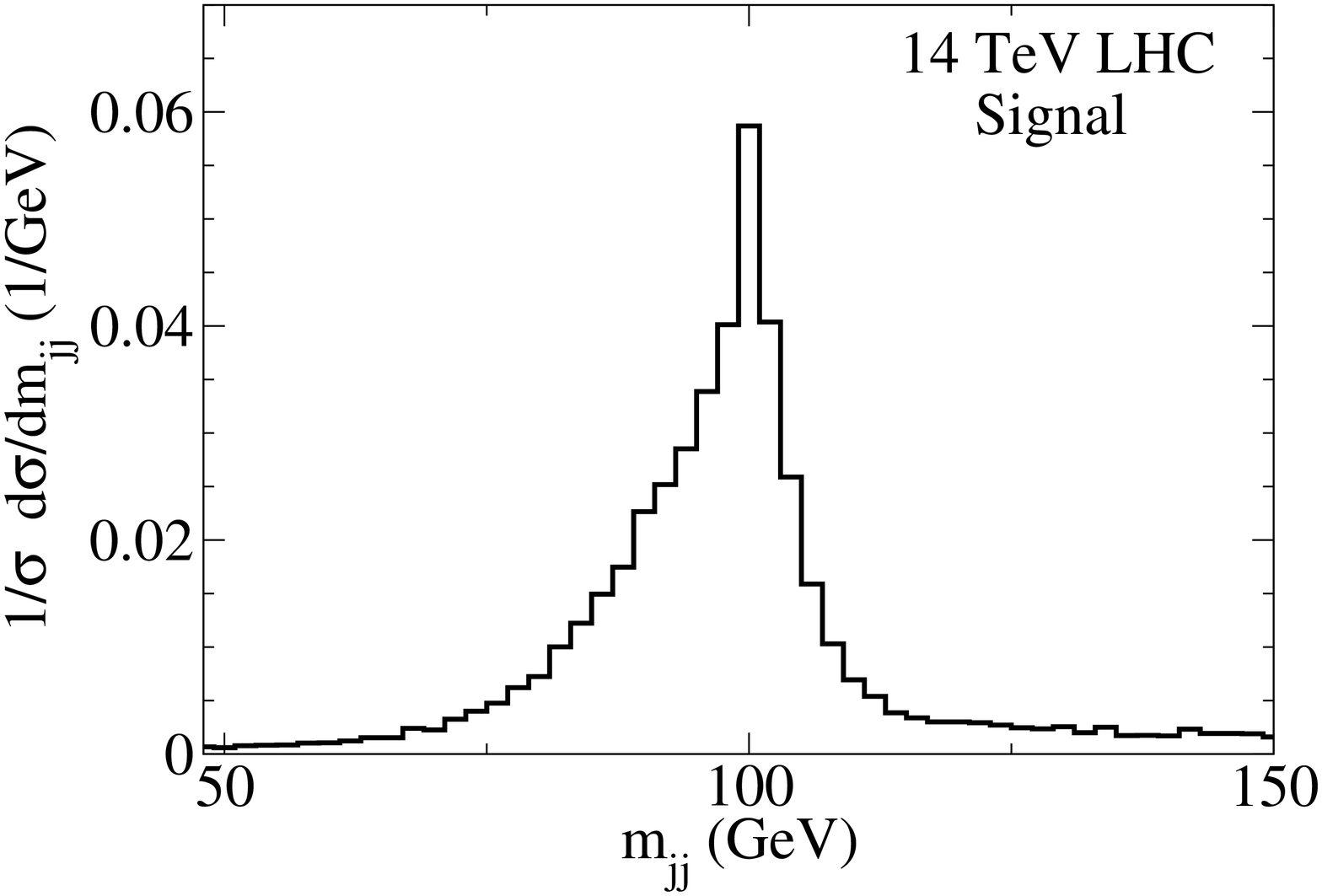}
	}
\subfigure[]{
	\includegraphics[width=0.31\textwidth,clip,angle=-90]{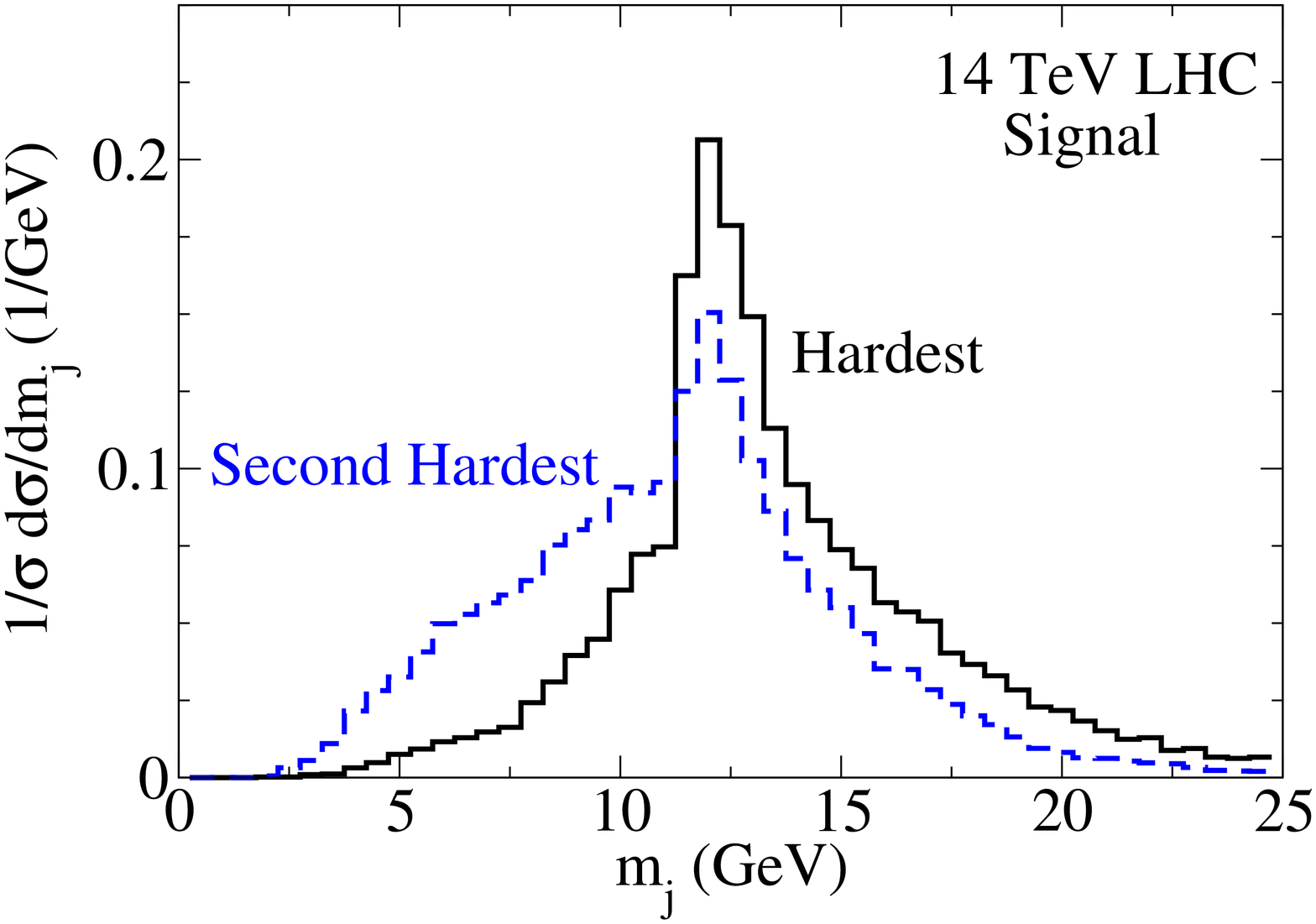}
	\label{mj.FIG}
}
\caption{ Invariant mass distributions of the signal $Wh\rightarrow \ell\nu 2j$ at the 14 TeV LHC for  (a) the two hardest jets and (b) the individual jet invariant mass for the hardest (solid) and second hardest (dashed) jets.  Cuts in Eq.~(3.1) and (3.2) have been applied.}
\label{mass.FIG}
\end{figure}

In Fig.~\ref{mass.FIG} we present the signal invariant mass distributions of (a) the dijet system consisting of the two hardest jets, $m_{jj}$, and (b) the individual jet masses, $m_j$, for the hardest (solid) and second hardest (dashed) jets.  The cuts in Eqs.~(\ref{cuts1J.EQ}) and (\ref{cuts1L.EQ}) have been applied.  As expected, the invariant mass of the dijets is peaked at the higgs mass $m_h\,=\,100$~GeV.  Also, since each jet originates from a massive particle, the individual jet masses peak at $m_j= m_\eta$.  The differences in the hardest and second hardest jet mass distributions can be understood by noting that higher order QCD processes can generate a jet mass.  For jets originating from massless partons, the average jet mass is approximately~\cite{arXiv:0712.2447}
\begin{eqnarray}
\langle m^2_j\rangle \approx C\frac{\alpha_s}{\pi}p^2_TR^2,
\label{mjpt.EQ}
\end{eqnarray}
where $R$ is the jet radius and $C$ depends on the initial partons.  Hence, jets with lower $p_T$ have on average lower masses than higher $p_T$ jets.  This effect can be seen in Fig.~\ref{mass.FIG}(b), where the second hardest jet is more (less) likely than the hardest jet to have a jet mass below (above) $m_\eta\,=\,12$~GeV.  Based on these observations we apply the invariant mass cuts 
\begin{eqnarray}
90~{\rm GeV}\,<\,m_{jj}\,<\,110~{\rm GeV}~~{\rm and}~~8~{\rm GeV}\,<\,m_{j}\,<\,16~{\rm GeV}.
\label{cuts2.EQ}
\end{eqnarray}
The effect of the invariant mass cuts on the signal is shown in the second column of Table~\ref{xsect.TAB}.

The irreducible backgrounds are the QCD background $Wjj$ and the electroweak backgrounds $WW/WZ\rightarrow \ell\nu 2j$.  Other contributing reducible backgrounds are $tq\rightarrow \ell\nu bj$, $tW\rightarrow \ell\nu b 2j$, $t\bar{t}\rightarrow \ell\nu b\bar{b} 2j$, and $tbW\rightarrow \ell\nu2b2j$.  The effects of the cuts in Eqs.~(\ref{cuts1J.EQ}), (\ref{cuts1L.EQ}), and (\ref{cuts2.EQ}) on the background cross sections are shown in the second and third columns of Table~\ref{xsect.TAB}.  Note that the $t\bar{t}$ processes are included in the $tbW$ background and are shown only for reference. As can be seen, the invariant mass cuts greatly reduce all backgrounds.  To further reduce the irreducible backgrounds, we require that the number of jets, $n^{\rm pass}_j$, passing the cuts in Eq.~(\ref{cuts1J.EQ}) to be equal to two.  The effects of the $n^{\rm pass}_j=2$ cut  on signal and background are shown in the fourth column of Table~\ref{xsect.TAB}.

\begin{figure}[tb]
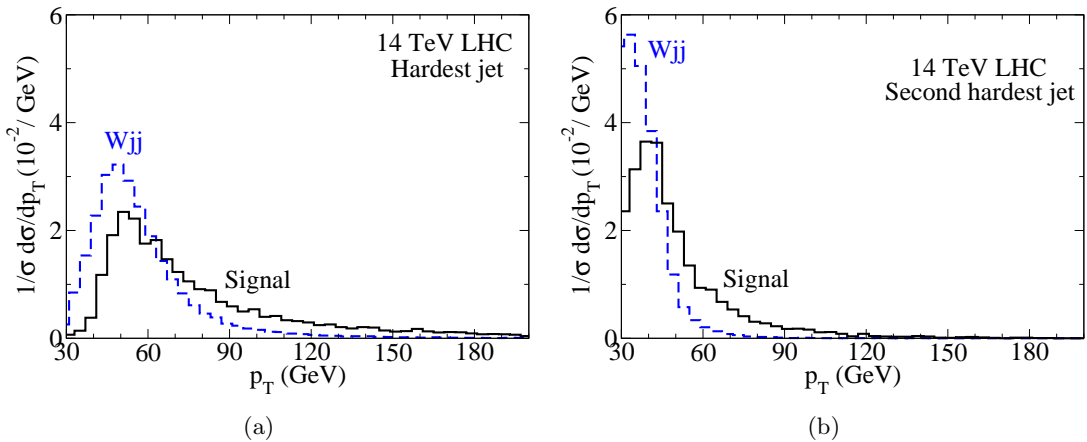

\centering
\subfigure[]{
	\label{ptWh.FIG}
	\includegraphics[width=0.45\textwidth,clip]{ptj_hard.eps}
	}
\subfigure[]{
	\includegraphics[width=0.45\textwidth,clip]{ptj_soft.eps}
	\label{ptWjj.FIG}
}
\caption{Transverse momentum distribution of the two jets passing the cuts in Eqs.~(3.1), (3.2), and (3.4) for (a) hardest and (b) second hardest jets at the 14 TeV LHC for both the $Wh$ signal (solid) and $Wjj$ background (dashed). }
\label{pT.FIG}
\end{figure}

After the previous cuts, $Wjj$ is still the dominant background.  Equation~(\ref{mjpt.EQ}) implies that the effect of the jet mass cut in Eq.~(\ref{cuts2.EQ}) is to cause the QCD jets to strongly peak at low $p_T$ with a short tail into the the high $p_T$ region.  Whereas, since the signal jets originate from massive particles, the jet mass cut is expected have much less effect on the signal jet $p_T$ distributions.  Hence, the signal jets are expected to have longer tails into the high $p_T$ region than the background jets.  In Fig.~\ref{pT.FIG} we show the transverse momentum distribution of the (a) hardest and (b) second hardest jets that pass the cuts in Eqs.~(\ref{cuts1J.EQ}), (\ref{cuts1L.EQ}), and (\ref{cuts2.EQ}) for both our $Wh$ signal (solid) and $Wjj$ background (dashed).    
As expected,  the background jets peak at lower $p_T$ and have shorter tails into the high $p_T$ region than the respective signal jets.  Hence, we apply the further $p_T$ cuts on signal and background jets:  
\begin{eqnarray}
p^j_T({\rm hard})\,>\,~100~{\rm GeV}~~{\rm and}~~p^j_T({\rm second~hardest})\,>\,~50~{\rm GeV}
\label{cuts3.EQ}
\end{eqnarray} 
The effect of these cuts on signal and background are shown in the fifth column of Table~\ref{xsect.TAB}.

After all the above cuts, the signal is still difficult to observe over the background.  With a luminosity of $30$ fb$^{-1}$ the significance at the $14$ TeV LHC is still only $0.78\sigma$, improving only slightly to $1.4\sigma$ at $100$ fb$^{-1}$.  As mentioned previously, our signal jets originate from color singlet massive particles while the background jets originate from colored massless partons.  We now exploit those differences and perform a subjet analysis to increase the signal significance.

\subsection{Jet Substructure}
\label{Subjet.SEC}
  For the subjet analysis we follow the standard procedure given in Ref.~\cite{Butterworth:2008iy}.  First, for the two jets passing all previous cuts, the final step in the jet reconstruction is reversed to determine the two leading subjets: $j_1$ and $j_2$ with $m_{j_1}>m_{j_2}$.  Since the parent jet originates from a massive particle and the subjets from massless partons, there is expected to be a significant mass drop between the subjets and parent jet.  Hence the subjet masses are required to meet the criteria $m_{j_1}<\mu\,m_j$, where $m_j$ is the parent jet's mass and $\mu$ is a free parameter.  Also, for the decay $\eta\rightarrow c\bar{c}$ the splitting should not to be too asymmetric
\begin{eqnarray}
\frac{\min({p^{j_1}_{T}}^2,{p^{j_2}_{T}}^2)}{m^2_j}\Delta R^2_{j_1,j_2}>y_{cut},
\end{eqnarray}
where $\Delta R^2_{j_1,j_2}=(\phi^{j_1}-\phi^{j_2})^2+(y^{j_1}-y^{j_2})^2$ measures the separation of the subjets in the $y-\phi$ plane and $\phi^{j_i}$ is the azimuthal angle of subjet $j_i$.  If the subjets do not satisfy the mass drop and asymmetric splitting criteria, the parent jet is replaced by $j_1$ and the procedure is repeated.  As noted in Ref.~\cite{Butterworth:2008iy}, for $\mu\gtrsim 1/\sqrt{3}$ the three-body decay $\eta\rightarrow c\bar{c}g$ will pass the mass drop criterion if the decay is in the Mercedes configuration in the $\eta$ rest frame.  Following   Refs.~\cite{Chen:2010wk,Butterworth:2008iy} we take $\mu=0.67$ and $y_{cut}=0.12$.

\begin{figure}[tb]
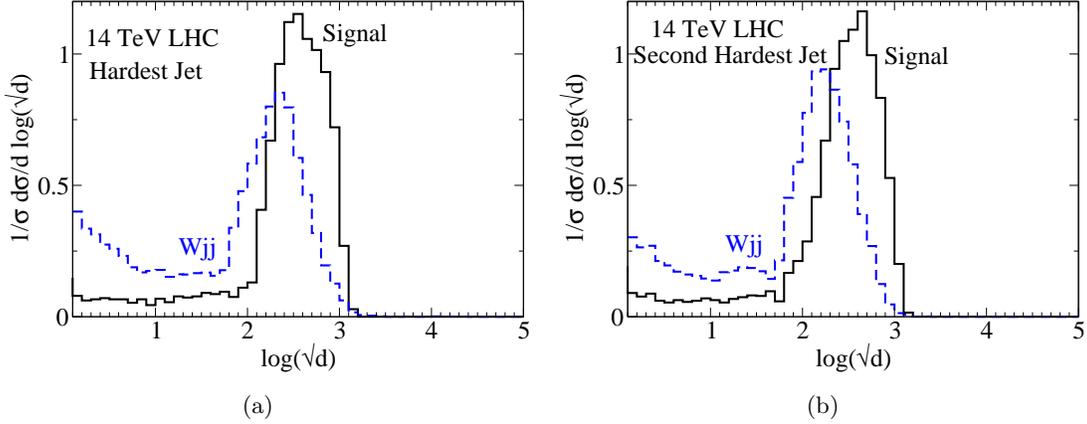

\centering
\subfigure[]{
	\label{sig_logkt.FIG}
	\includegraphics[width=0.45\textwidth,clip]{logkt_hard.eps}
	}
\subfigure[]{
	\includegraphics[width=0.45\textwidth,clip]{logkt_soft.eps}
	\label{wjj_logkt.FIG}
}
\caption{ $\log\sqrt{d}$ distributions at the 14 TeV LHC for  (a) the hardest jet and (b) second hardest jet for both $Wh$ signal (solid) and $Wjj$ background (dashed).  Cuts in Eq.~(3.1), (3.2), and (3.4) have been applied and $d$ is measured in units of GeV$^2$.}
\label{KT.FIG}
\end{figure}

Each jet in our signal is expected to result from the $\eta\rightarrow2c$ decay, where the two quarks will be collimated into a single jet.  Hence the KT distance of the two subjets $d=\min({p^{j_1}_T}^2,{p^{j_2}_T}^2)\Delta R^2_{j_1,j_2}/R^2\sim \mathcal{O}(m^2_\eta)$~\cite{Butterworth:2002tt}.  Figure~\ref{KT.FIG} shows the $\log\sqrt{d}$ distributions for the (a) hardest and (b) second hardest jets with $d$ measured in GeV$^2$ and after the cuts in Eqs.~(\ref{cuts1J.EQ}),~(\ref{cuts1L.EQ}), and (\ref{cuts2.EQ}) are applied.  The $Wh$ signal distribution is shown with solid lines and the $Wjj$ background with dashed.  Since we apply the jet mass cut in Eq.~(\ref{cuts2.EQ}),  both the signal and background distributions are peaked near the same value of $\log\sqrt{d}$.  However, since the signal jets have a natural mass scale of $m_\eta=12$~GeV, the signal distributions are peaked narrowly at $\log\sqrt{d}\sim\log 12\approx 2.5$.  Also, since the $Wjj$ jets originate from massless partons, the background distribution has a significant tail in the low $\log\sqrt{d}$ region and is peaked at a slightly lower value than $2.5$.  We therefore apply the subjet cuts:
\begin{eqnarray}
2\,<\,\log\sqrt{d}\,<3.
\label{cuts5.EQ}
\end{eqnarray}

\begin{figure}[tb]
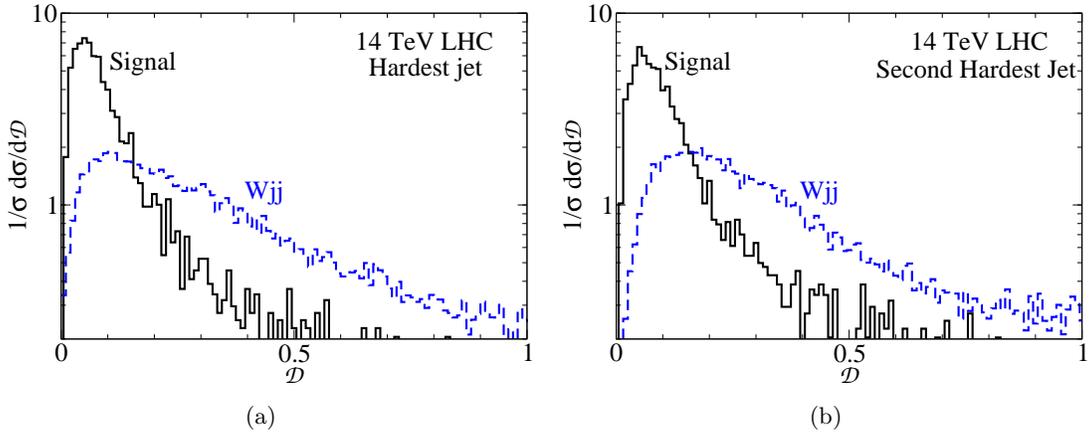

\centering
\subfigure[]{
	\label{sipol_hard.FIG}
	\includegraphics[width=0.45\textwidth,clip]{dipol_hard.eps}
	}
\subfigure[]{
	\includegraphics[width=0.45\textwidth,clip]{dipol_soft.eps}
	\label{dipol_soft.FIG}
}
\caption{ Dipolarity distributions of the signal $Wh\rightarrow \ell\nu 2j$ at the 14 TeV LHC for  (a) the hardest jet and (b) second hardest jet for both $Wh$ signal (solid) and background (dashed).  Cuts in Eqs.~(3.1), (3.2), and (3.4) have been applied.}
\label{dipol.FIG}
\end{figure}

Since our signal jets originate from color singlet particles, the two subjets within each jet are expected to form a color dipole.   Hence, for our signal most of the radiation is expected to lie between the two subjets; whereas, the radiation in the background jets is expected to be more uniformly distributed since the jets originate from colored partons.  The observable ``dipolarity"~\cite{Hook:2011dp} has been proposed to measure whether or not subjets form a color dipole:
\begin{eqnarray}
\mathcal{D}=\frac{1}{\Delta R^2_{j_1,j_2}}\sum_{i\epsilon J}\frac{p_{T_i}}{p_{T_J}}R^2_i
\end{eqnarray}
where the sum runs over all calorimeter cells within a jet, $p_{T_i}$ is the $p_T$ in a given calorimeter cell, and $p_{T_J}$ is the jet's total $p_T$.  The distance between a given calorimeter cell and the line connecting the centers of the two subjets in the $y-\phi$ plane is given by $R_i$.  Fig.~\ref{dipol.FIG} shows the dipolarity of the (a) hardest and (b) second hardest jets after cuts in Eq.~(\ref{cuts1J.EQ}), (\ref{cuts1L.EQ}), and (\ref{cuts2.EQ}). Again, the $Wh$ signal is shown with solid lines and the $Wjj$ background with dashed lines. For this distribution the dipolarity has been calculated at the hadronic level. As can be clearly seen, the distribution for the signal jets is peaked at low dipolarity and the $Wjj$ distributions are more uniformly spread.  We therefore apply the dipolarity cut
\begin{eqnarray}
\mathcal{D}\,<\,~0.1
\label{cuts6.EQ}
\end{eqnarray}

The effects of the cuts of Eqs.~(\ref{cuts5.EQ}) and (\ref{cuts6.EQ}) are shown in the last column of Table~\ref{xsect.TAB}. Including the subjet cuts improves the statistical significance of the signal to $3.8\sigma$ at $30$ fb$^{-1}$ and $7.0\sigma$ at 100 fb$^{-1}$ of integrated luminosity at the 14 TeV LHC.  A statistical significance of $5\sigma$ can be obtained with approximately $50$ fb$^{-1}$ of data.

\section{Conclusions}
\label{Conc.SEC}
The LHC is quickly excluding the remaining parameter ranges for the SM Higgs boson.  Together with LEP direct detection limits, the only remaining available masses for a low mass SM Higgs are $117.5-118.5$~GeV and $122.5-129$~GeV~\cite{Barate:2003sz,ATLAS,CMS}.  As these limits tighten, we are motivated to investigate other mechanisms of EWSB.  Particularly interesting models are those that escape the LEP bounds and can have a Higgs boson with a mass around $100$~GeV.  This can be accomplished if the Higgs boson undergoes a cascade decay $h\rightarrow 2\eta\rightarrow 4X$, where $\eta$ is a pseudoscalar and $X$ are light SM particles~\cite{Dermisek:2005ar}.  

We have studied the observability of a Higgs boson in the ``charming Higgs" model~\cite{Bellazzini:2009kw}.  In this model the Higgs decays via $h\rightarrow2\eta\rightarrow4c$ and can escape the LEP bounds.  In particular, we considered the associated production of a Higgs boson with leptonically decaying $W$ for masses $m_h=100$ GeV and $m_\eta=12$ GeV. Since $m_h\gg m_\eta$, the decay products of $\eta\rightarrow c\bar{c}$ are highly collimated and form single jets.  Our signal then consists of a charged lepton, missing energy, and two jets originating from the two $\eta$s from the Higgs decay. Such a signal is typically overwhelmed by the $Wjj$ QCD background.  By applying cuts to the invariant mass and transverse momentum of the jets, as well as requiring that exactly two jets satisfy our basic acceptance cuts, we can partially lift this signal above the standard model background at the LHC, but still only reach $S/\sqrt{S+B}$ of  $0.78\sigma$  at $30$ fb$^{-1}$ and $1.4\sigma$ at $100$ fb$^{-1}$ at the $14$ TeV LHC.

Our analysis can be greatly improved by applying jet substructure techniques. We require a significant mass drop from jet to subjet and fairly symmetric splitting of momentum between the two subjets. Additionally, we place new cuts on the KT distance between our subjets (which should be peaked around $m_{\eta}^2$ for our signal) and the jet dipolarity~\cite{Hook:2011dp} (because our  signal comes from color singlets whereas our background is dominated by colored particles). Including these subjet cuts allows us to achieve a statistical significance of almost $4\sigma$ at $30$ fb$^{-1}$ and $7\sigma$ at $100$ fb$^{-1}$ at the $14$~TeV LHC.  A $5\sigma$ discovery can be achieved with $50$ fb$^{-1}$ of data.

The procedures presented here are quite general and applicable to situations in which the Higgs boson undergoes the decay $h\rightarrow 2\eta\rightarrow 4j$ with $m_\eta\ll m_h$.  Also, the branching ratio of $h\rightarrow 4c$ may differ from one and the Higgs mass may be lager than $100$~GeV.  Simply scaling the signal rate with branching ratio, we estimate that with $100$~fb$^{-1}$ of data at the $14$ TeV LHC, branching ratios of BR$(h\rightarrow 2\eta\rightarrow 4c)=0.35$ and $0.64$ are observable at the $3\sigma$ and $5\sigma$ levels, respectively.  Also, as the Higgs mass increases the dijet invariant mass cuts will need to be tuned and the signal rate will decrease slightly due to the higher final state invariant mass.  However, the favored signal dijet mass is further from the  $Z$-pole and the dijet invariant mass cuts will be more efficient in reducing the $WW$ and $WZ$ backgrounds. With the tuned dijet invariant mass cuts, the QCD background is expected to decrease at least as quickly as the signal rate.  Hence, a slightly higher Higgs mass is not expected to significantly alter our conclusions.

\section{Acknowledgement}
We would like to thank Prof. Tao Han for suggesting this project and for helpful discussions.  This work was supported in part by  the U.S.~DOE under Grant
No.~DE-FG02-95ER40896. IL was also supported in part by the US~DOE Grant No.~DE-AC02-98CH10886.


\begin{thebibliography}{99}


\bibitem{Barate:2003sz}
  R.~Barate {\it et al.} [ LEP Working Group for Higgs boson searches and ALEPH and DELPHI and L3 and OPAL Collaborations ],
  Phys.\ Lett.\  {\bf B565}, 61-75 (2003).
  [hep-ex/0306033].

\bibitem{ATLAS}
 ATLAS Collaboration,  ATLAS-CONF-2012-019
\bibitem{CMS}
 CMS Collaboration, CMS PAS HIG-12-008

\bibitem{Weinberg:1975gm}
  S.~Weinberg,
  Phys.\ Rev.\  {\bf D13}, 974-996 (1976);
  L.~Susskind,
  Phys.\ Rev.\  {\bf D20}, 2619-2625 (1979).

\bibitem{:2010zz}
  ALEPH, CDF, D0, DELPHI, L3, OPAL, SLD, the LEP Electroweak Working Group, the Tevatron Electroweak Working Group, and the SLD electroweak and heavy flavour groups Collaboration, arXiv:1012.2367 [hep-ex] 


\bibitem{Dermisek:2005ar}
  R.~Dermisek, J.~F.~Gunion,
  Phys.\ Rev.\ Lett.\  {\bf 95}, 041801 (2005).
  [hep-ph/0502105];
  J.~F.~Gunion,
  Int.\ J.\ Mod.\ Phys.\ A {\bf 25}, 4163 (2010).


\bibitem{Dobrescu:2000yn} 
  B.~A.~Dobrescu and K.~T.~Matchev,
  JHEP {\bf 0009}, 031 (2000)
  [hep-ph/0008192];
  M.~S.~Carena, J.~R.~Ellis, S.~Mrenna, A.~Pilaftsis and C.~E.~M.~Wagner,
  Nucl.\ Phys.\ B {\bf 659}, 145 (2003)
  [hep-ph/0211467].


\bibitem{Chang:2006bw}
  B.~A.~Dobrescu, G.~L.~Landsberg, K.~T.~Matchev,
  Phys.\ Rev.\  {\bf D63}, 075003 (2001).
  [hep-ph/0005308];
  S.~Chang, P.~J.~Fox, N.~Weiner,
  Phys.\ Rev.\ Lett.\  {\bf 98}, 111802 (2007).
  [hep-ph/0608310];
  K.~Cheung, J.~Song, Q.~-S.~Yan,
  Phys.\ Rev.\ Lett.\  {\bf 99}, 031801 (2007).
  [hep-ph/0703149];
  S.~Chang, R.~Dermisek, J.~F.~Gunion, N.~Weiner,
  Ann.\ Rev.\ Nucl.\ Part.\ Sci.\  {\bf 58}, 75-98 (2008).
  [arXiv:0801.4554 [hep-ph]];
  M.~Carena, T.~Han, G.~-Y.~Huang, C.~E.~M.~Wagner,
  JHEP {\bf 0804}, 092 (2008).
  [arXiv:0712.2466 [hep-ph]];
  M.~Lisanti, J.~G.~Wacker,
  Phys.\ Rev.\  {\bf D79}, 115006 (2009).
  [arXiv:0903.1377 [hep-ph]];
  A.~Belyaev, J.~Pivarski, A.~Safonov, S.~Senkin, A.~Tatarinov,
  Phys.\ Rev.\  {\bf D81}, 075021 (2010).
  [arXiv:1002.1956 [hep-ph]].
  B.~Bellazzini, C.~Csaki, J.~Hubisz, J.~Shao,
  Phys.\ Rev.\  {\bf D83}, 095018 (2011).
  [arXiv:1012.1316 [hep-ph]];
 C.~Englert, T.~S.~Roy and M.~Spannowsky,
 Phys.\ Rev.\ D {\bf 84}, 075026 (2011)
 [arXiv:1106.4545 [hep-ph]].

\bibitem{Falkowski:2010hi}
  A.~Falkowski, D.~Krohn, L.~-T.~Wang, J.~Shelton, A.~Thalapillil,
  [arXiv:1006.1650 [hep-ph]];
  D.~E.~Kaplan, M.~McEvoy,
  Phys.\ Rev.\  {\bf D83}, 115004 (2011).
  [arXiv:1102.0704 [hep-ph]].

\bibitem{Chen:2010wk}
  C.~-R.~Chen, M.~M.~Nojiri, W.~Sreethawong,
  JHEP {\bf 1011}, 012 (2010).
  [arXiv:1006.1151 [hep-ph]];

\bibitem{Schael:2006cr}
  S.~Schael {\it et al.} [ ALEPH and DELPHI and L3 and OPAL and LEP Working Group for Higgs Boson Searches Collaborations ],
  Eur.\ Phys.\ J.\  {\bf C47}, 547-587 (2006).
  [hep-ex/0602042].

\bibitem{Bellazzini:2009kw}
  B.~Bellazzini, C.~Csaki, A.~Falkowski, A.~Weiler,
  Phys.\ Rev.\  {\bf D81}, 075017 (2010).
  [arXiv:0910.3210 [hep-ph]].

\bibitem{Butterworth:2008iy}
  J.~M.~Butterworth, A.~R.~Davidson, M.~Rubin, G.~P.~Salam,
  Phys.\ Rev.\ Lett.\  {\bf 100}, 242001 (2008).
  [arXiv:0802.2470 [hep-ph]].

\bibitem{Bellazzini:2009xt}
  B.~Bellazzini, C.~Csaki, A.~Falkowski, A.~Weiler,
  Phys.\ Rev.\  {\bf D80}, 075008 (2009).
  [arXiv:0906.3026 [hep-ph]].


\bibitem{Alwall:2011uj}
  J.~Alwall, M.~Herquet, F.~Maltoni, O.~Mattelaer, T.~Stelzer,
  JHEP {\bf 1106}, 128 (2011).
  [arXiv:1106.0522 [hep-ph]].

\bibitem{Corcella:2000bw}
  G.~Corcella, I.~G.~Knowles, G.~Marchesini, S.~Moretti, K.~Odagiri, P.~Richardson, M.~H.~Seymour, B.~R.~Webber,
  JHEP {\bf 0101}, 010 (2001).
  [arXiv:hep-ph/0011363]; [arXiv:hep-ph/0210213]

\bibitem{Christensen:2008py}
  N.~D.~Christensen, C.~Duhr,
  Comput.\ Phys.\ Commun.\  {\bf 180}, 1614-1641 (2009).
  [arXiv:0806.4194 [hep-ph]].

\bibitem{Cacciari:2006sm}
  M.~Cacciari,
  [hep-ph/0607071], http://fastjet.fr

\bibitem{Catani:1993hr}
  S.~Catani, Y.~L.~Dokshitzer, M.~H.~Seymour, B.~R.~Webber,
  Nucl.\ Phys.\  {\bf B406}, 187-224 (1993);
  S.~D.~Ellis, D.~E.~Soper,
  Phys.\ Rev.\  {\bf D48}, 3160-3166 (1993).
  [hep-ph/9305266].


\bibitem{arXiv:0712.2447} 
  S.~D.~Ellis, J.~Huston, K.~Hatakeyama, P.~Loch and M.~Tonnesmann,
  Prog.\ Part.\ Nucl.\ Phys.\ \ {\bf 60}, 484  (2008)
  [arXiv:0712.2447 [hep-ph]].

\bibitem{Butterworth:2002tt} 
  J.~M.~Butterworth, B.~E.~Cox and J.~R.~Forshaw,
  Phys.\ Rev.\ D {\bf 65}, 096014 (2002)
  [hep-ph/0201098];
  G.~Aad {\it et al.}  [The ATLAS Collaboration],
  arXiv:0901.0512 [hep-ex];
  J.~M.~Butterworth, J.~R.~Ellis and A.~R.~Raklev,
  JHEP {\bf 0705}, 033 (2007)
  [hep-ph/0702150 [HEP-PH]].

 
\bibitem{Hook:2011dp}
  A.~Hook, M.~Jankowiak and J.~G.~Wacker,
  JHEP {\bf 1204}, 007 (2012)
  [arXiv:1102.1012 [hep-ph]].


\end{thebibliography}
\end{document}